\RequirePackage{amsmath}
\documentclass{llncs}
\usepackage{times}
\usepackage{helvet}
\usepackage{courier}
\usepackage{hyperref}
\usepackage{graphicx}
\usepackage{amsmath}
\usepackage[round]{natbib}
\usepackage[affil-it]{authblk} % Include the authblk package
\begin{document}
\title{EEGMobile: Enhancing Speed and Accuracy in EEG-Based Gaze Prediction with Advanced Mobile Architectures}

\author{Teng Liang\inst{1}\orcidID{0009-0009-3074-2420}
 \and 
Andrews Damoah\inst{2}\orcidID{0009-0006-2840-6063}}  %orcid: 

\institute{Palo Alto High School \\
\email{tl40355@pausd.us} \and 
University of Maryland - College Park \\
\email{adamoah@terpmail.umd.edu}}

%Remove date
\date{}
\maketitle
\thispagestyle{plain}

\begin{abstract}
Electroencephalography (EEG) analysis is an important domain in the realm of Brain-Computer Interface (BCI) research. To ensure BCI devices are capable of providing practical applications in the real world, brain signal processing techniques must be fast, accurate, and resource-conscious to deliver low-latency neural analytics. This study presents a model that leverages a pre-trained MobileViT alongside Knowledge Distillation (KD) for EEG regression tasks. Our results showcase that this model is capable of performing at a level comparable (only 3\% lower) to the previous State-Of-The-Art (SOTA) on the EEGEyeNet Absolute Position Task while being 33\% faster and 60\% smaller. Our research presents a cost-effective model applicable to resource-constrained devices and contributes to expanding future research on lightweight, mobile-friendly models for EEG regression.

\end{abstract}

%too long make shorter
%overall add more about the contribution to HCI
\section{1. Introduction}
Electroencephalography (EEG) signal analysis is a pivotal research subject contributing to the advancement of Brain-Computer Interfaces (BCI). Furthermore, the application of Machine Learning (ML) and Deep Learning (DL) algorithms has become a key component for EEG analysis, which has only grown steadily across the years \citep{eegdirection}. The EEGEyeNet dataset has become the centerpiece in this field fusing advanced EEG data compilation with cutting-edge ML and DL techniques. \citep{murungi2023trends,dou2022time,wolf2022deep,rolff2022gazetransformer,kastrati2023electrode,farago2022blink}. Its popularity is underscored by its large collection of high-quality EEG and eye-tracking (ET) recordings alongside baseline ML and DL models to benchmark accuracy on a variety of ET tasks. The advent of EEGEyeNet has spawned further research into applications of new DL algorithms for EEG-ET tasks \cite{vit2eeg, modesitt2024fusing, dou2022time, rolff2022gazetransformer}. While much of this research is focused on improving predictive accuracy, for these methods to be practical in real-world scenarios, there is a need to develop more computationally efficient models.

As such, this study explores the efficacy of integrating a pre-trained MobileViT network into the existing EEG-based transformer model (EEGViT-TCNet) alongside incorporating knowledge distillation during training with a fine-tuned EEGViT-TCNet teacher model for EEG-based gaze estimation. This unique approach allows our model, dubbed EEGMobile, to leverage the MobileViT's effective design to increase computational efficiency during inferencing. The use of knowledge distillation also enhances task performance, allowing our model to exhibit vastly superior accuracy when compared to previous iterations where knowledge distillation was not used.

This research provides an extensive evaluation of EEGMobile and contrasts it with previous models focused on the same task, both in terms of speed and accuracy. We aim to highlight the strengths of the individual components that comprise our model as a whole and illustrate how these components unite to form a cost-effective model for EEG regression. Our research has the potential to expand the efficiency and scalability of practical applications of DL models for EEG analysis across a plethora of domains. Our code is publically available at: \href{https://github.com/t0nyliang/EEGMobile}{https://github.com/t0nyliang/EEGMobile}.

\subsection{1.1 Research Question}
\begin{itemize}
    \item Is a MobileViT-based model viable for faster, SOTA-comparable accuracy on the EEGEyeNet dataset?
\end{itemize}
%add a couple sentences below on the contribution to HCI
By answering this question, we hope to contribute to research into the development of more memory-conscious models designed for resource-constrained devices, such as mobile phones. Our research contributions highlight the potential for expanding the accessibility of EEG-based eye-tracking technology to a wider audience, further interpolating neuroscience and Human-Computer Interaction (HCI) for medical applications.

%Acronym table here?
%shrink the size of the table after the abstract is in
\begin{table}[htbp]
\caption{\textit{Abbreviation Table}}
\label{tab:abbreviation}
\centering
\resizebox{0.6\textwidth}{!}{%
\begin{tabular}{|l|c|c|}
\hline
\textbf{Abbreviation} & \textbf{Definition} \\ \hline
EEG & Electroencephalography \\ \hline
ET & Eye-Tracking \\ \hline
BCI & Brain-Computer Interfaces \\ \hline
AI & Artificial Intelligence \\ \hline
ML & Machine Learning \\ \hline
DL & Deep Learning \\ \hline
ViT & Vision Transformer \\ \hline
CNN & Convolutional Neural Network \\ \hline
TCN & Temporal Convolutional Network \\ \hline
KD & Knowledge Distillation \\ \hline
SOTA & State Of The Art \\ \hline
RMSE & Root Mean Squared Error \\ \hline
\end{tabular}
}
\end{table}

%EEGEyeNet
%EEGVit and EEGVit-TCN papers (possible just tcn or both)
%MobileVit 1 and 2 (https://arxiv.org/abs/2110.02178 | https://arxiv.org/abs/2206.02680)
%Knowledge distillation (https://arxiv.org/abs/1503.02531) this is the original but I will see if there is something more recent and relevant
\section{2. Related Work} %ADD CITATIONS!!!!
%Make sure to differentiate the discussion of the EEGEyeNet paper in this section from the discussion of the dataset in the methods section to prevent repetition

\begin{table}[hbtp]
\caption{\textit{Root Mean Squared Error (RMSE) of EEGEyeNet Random guessing and baseline DL models, EEGViT, and EEGViT-TCN \citep{eegeyenet,vit2eeg}}}
\label{tab:model_comparison}
\centering
\resizebox{0.94\textwidth}{!}{%
\begin{tabular}{|l|c|}
\hline
\textbf{Model} & \textbf{Absolute Position RMSE (mm)} \\ \hline
Naive Guessing & 123.3 ± 0.0 \\ \hline \hline
CNN & 70.4 ± 1.1 \\ \hline
PyramidalCNN & 73.9 ± 1.9 \\ \hline
EEGNet & 81.3 ± 1.0 \\ \hline
InceptionTime & 70.7 ± 0.8 \\ \hline
Xception & 78.7 ± 1.6 \\  \hline \hline
ViT-Base & 58.1 ± 0.6 \\ \hline
ViT-Base & 61.5 ± 0.6 \\ \hline
ViT-Base (Pre-trained) & 58.1 ± 0.6 \\ \hline
EEGViT & 61.7 ± 0.6 \\ \hline
EEGViT (Pre-trained) & 55.4 ± 0.2 \\ \hline \hline
EEGViT-TCN & 51.8 ± 0.2 \\ \hline

\end{tabular}

}
\end{table}

\subsection{2.1 EEG and Deep Learning}

Electroencephalography (EEG) is extensively used in various research domains, including neural engineering, neuroscience, biomedical engineering, and brain-like computing, especially in the development of brain-computer interfaces (BCIs). The study of EEG signals is essential for the progress of BCIs, providing deep insights into the complex neural activities of the human brain.

In the past decade, a wide range of machine learning and deep learning algorithms have been applied to EEG data, leading to significant advancements in numerous applications. These applications include emotion recognition, motor imagery, mental workload assessment, seizure detection, Alzheimer's disease classification, and sleep stage scoring, among others \citep{craik2019deep,kastrati2021eegeyenet,roy2019deep,altaheri2023deep,qu2022time,gao2021complex,hossain2023status,yi2022attention,key2024advancing,li2024enhancing,koome2023trends,murungi2023empowering,dou2022time,zhou2022brainactivity1,qu2020identifying,qu2020using,qu2020multi,qu2018eeg,qu2019personalized,saeidi2021neural, qu2022eeg4home,rasheed2020machine, dadebayev2022eeg,wang2022eeg,li2020deep,aggarwal2022review}. EEG and deep learning research have been in close proximity for decades, with advancements in both fields contributing to strides in our understanding of the brain. Deep learning algorithms are particularly popular in the context of EEG analysis due to their ability to extrapolate and generalize input information, making them ideal for decoding the complexities and noise within EEG signals into interpretable outputs.

The research presented in the EEGEyeNet study encapsulates this main point, presenting a large-scale EEG and eye-tracking dataset, designed specifically for ML and DL research \cite{eegeyenet}. The study also presents a robust baseline for comparing the performance of new models on gaze estimation tasks. Due to its high-quality and accessible dataset, EEGEyeNet has been used by numerous studies experimenting with various DL techniques and architectures to improve performance. Their results, shown in Table \ref{tab:model_comparison}, specifically highlight Convolutional Neural Networks (CNNs) as having a profound ability to interpret complex EEG data accurately, achieving a Root Mean Square Error (RMSE) value of 70.4. 
The contributions of combined EEG and deep learning heavily deepen our understanding of the brain and thus influence several related fields. EEG-based Artificial Intelligence (AI) systems can allow for the automated detection and monitoring of various neurological states and conditions, advancing fields such as neuroscience and BCI. In the case of EEGEyeNet, behavioral and neurological information can be deduced from eye tracking, further highlighting the importance of enhancing the capabilities of deep learning algorithms for EEG regression \cite{cao2020eegai}.

%Mention that this is using the Absolute Position task, and what it actually is

\begin{figure}[hbtp]
  \centering
  \includegraphics[width=1.0 \linewidth]{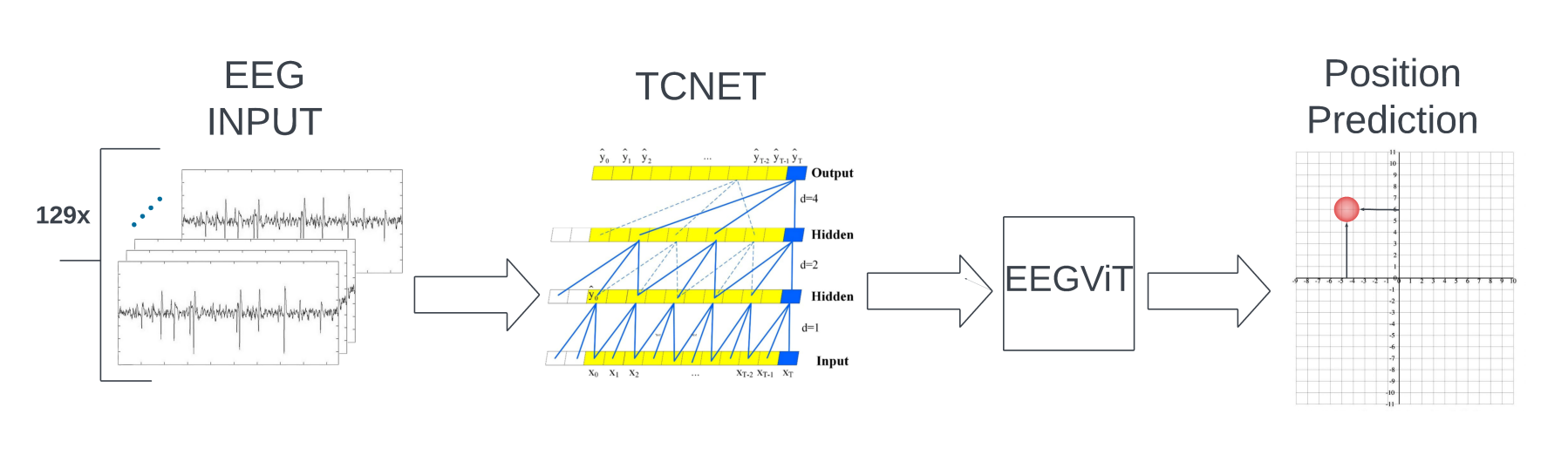}
  \caption{\textit{EEGViT-TCNet Model Diagram \citep{modesitt2024fusing}.}}
  \label{fig:tcn}
\end{figure}

\subsection{2.2 Vision Transformers for EEG Regression}
Vision Transformers are well regarded as being extremely adept at performing a wide variety of image tasks when pre-trained on large datasets \cite{dosovitskiy2021image}. This is a result of the self-attention mechanism, which captures information across sequences of pixel patches, allowing the model to build a more robust representation of the entire image. While ViTs have been nominally utilized in image analysis, recent research has unmasked the applicability of ViTs for EEG analysis. The study presenting the EEGViT model highlighted how transformers' strong global and sequential data processing performs better at regression on time-series EEG data \cite{vit2eeg}. Furthermore, the addition of prior feature extraction layers, such as a Temporal Convolutional Network (TCN) block shown in Figure \ref{fig:tcn}, as discussed in EEGViT-TCNet, further extracts temporal features within sequences of EEG signals \citep{eegtcn}.

The strength of pre-trained ViTs for EEG regression lies in their ability to interpret sequence data as a whole and apply image-related priors to the EEG space. Moreover, evidence seems to suggest that decreased local connectivity may lead to more subject and noise invariance, broadening the variety of signal patterns that can be processed. Overall, ViTs prove to be an exceptional tool in the field of EEG signal analysis, overcoming the many limitations of other DL algorithms specific to EEG (Table \ref{tab:model_comparison}) \cite{vit2eeg}.

%captions for tables are suppose to be above the table

\begin{figure}[hbtp]
  \centering
  \includegraphics[width=1.0 \linewidth]{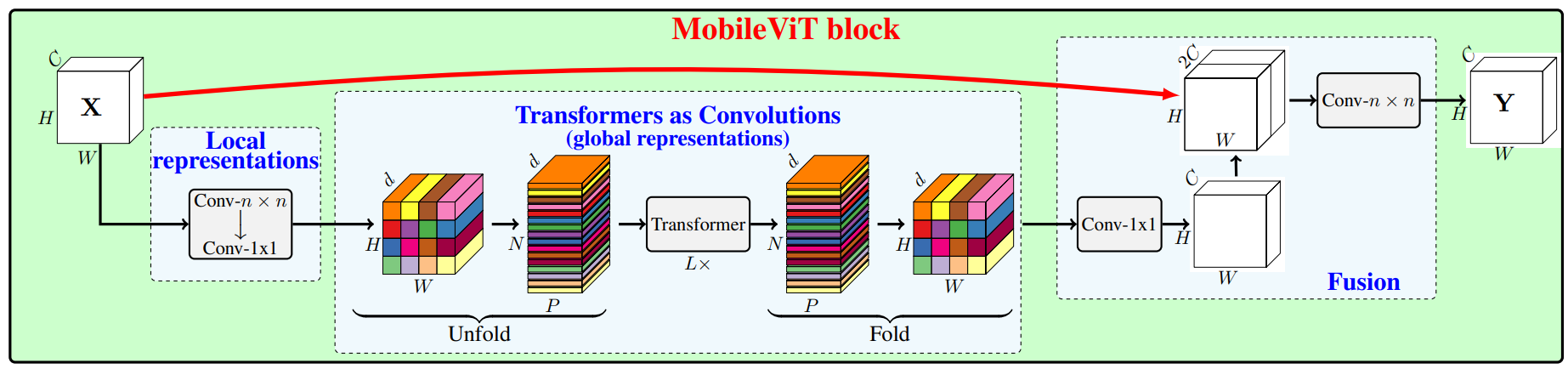}
  \caption{\textit{Model diagram for the MobileViT block, primary innovation of the MobileViT \citep{mobilevit}.}}
  \label{fig:mobile}
\end{figure}

\subsection{2.3 Lightweight and Mobile Networks}
The use of lightweight models is the standard for vision tasks on mobile or resource-constrained devices. Typically, these models are adapted from CNNs, which, due to their spatial inductive bias, yield better performance than ViTs at lower parameter sizes. Lightweight CNNs can also be more effortlessly adapted for downstream tasks and are generally easier to optimize than ViTs, which require extensive data augmentation and regularization \cite{xiao2021early}. However, to learn global representations, ViTs need to be integrated in some manner. MobileViT introduces a hybrid, lightweight model that treats transformers as convolutions \cite{mobilevit}. They achieve this by modifying standard convolution operations to encode global features via a transformer, seen in Figure \ref{fig:mobile}, permitting both local and global processing, resulting in better accuracy than other lightweight networks. Further optimizations were made to MobileViT, in the subsequently named MobileViTV2. This primarily involved altering the self-attention mechanism from using the multi-headed standard to a separable approach. Using this method, rather than attention scores being computed with respect to each patch, they are computed with respect to a single latent token, reducing the time complexity from quadratic to linear with respect to the number of patches. Additionally, the removal of the skip connection within the MobileViT block provided a minor improvement to the model's task performance.

The advancements of lightweight networks have implications directly related to building more deployable AI systems. Faster and more accurate models improve vision-related tasks on mobile devices, such as image recognition and augmented reality functionality. This extends outward from solely mobile phones and can apply to other resource-constrained devices, such as drones and wearable technology. When it comes to EEG analysis, both CNNs and Transformers are proven to be powerful tools for enhancing performance, and the adaptation of mobile networks can make strides in the practical application of DL in EEG signal processing \cite{eegeyenet, vit2eeg, modesitt2024fusing}.

\subsection{2.4 Knowledge Distillation for Lightweight Models}
While lightweight models perform reasonably well across various image task benchmarks, they still fall short in terms of accuracy when compared to their larger counterparts. Due to the necessity of a resource-conscious design, architectural modifications are more difficult to make, limiting their overall performance \cite{gao2023sisr}. One proven way to improve the performance of lightweight models is through model compression, with knowledge distillation being one well-known technique to achieve this. The main methodology behind this procedure is to have a larger, more complex network (teacher) with a smaller, simpler model (student) and train the student model using a "distillation loss" calculated from the combined losses of the student and teacher models \cite{knowledgedistilling}. This causes the student model to mimic the teacher model's behavior, effectively "compressing" the teacher into a smaller network. This is especially good for lightweight networks, as they are able to enjoy enhanced predictive accuracy alongside their computational efficiency.

Knowledge distillation is an incredibly popular and useful technique in designing deployable DL models for resource-constrained devices, as this training procedure can result in better accuracy without affecting the model's size or speed \cite{cui2024lightweight}. This makes it an attractive method to aid in developing enhanced lightweight networks for EEG analysis. These networks, with higher accuracy and efficiency, have the potential to increase practical applications in medical contexts, where speed and accuracy are vital.
%knowledge distillation

\section{3. Methods}

In our study, we concentrated on the Absolute Position task within the EEGEyeNet dataset, opting for the MobileViT architecture due to its proven superior performance relative to other similarly sized models. We incorporated a meticulously fine-tuned EEGViT-TCNet as the teacher model in our knowledge distillation process, aimed at enhancing the MobileViT student model's accuracy. The accuracy of our proposed model was rigorously evaluated using the Root Mean Square Error (RMSE) on the test set, in addition to measuring the computational speed during inference, and the model's parameter count. 

\subsection{3.1 Dataset}
This study employed the EEGEyeNet dataset for the training and validation of our model. \citep{wang2022cnn,fuhl2023one,xiang2022vector,modesitt2023two,mishra2023signeeg} This dataset encompasses EEG and ET recordings from 356 adults, 190 of whom were female and 166 were male, ranging in age from 18 to 80 years old. Researchers obtained written consent from all participants prior to data collection and compensated participants monetarily. The EEG data was recorded on the EEG Geodesic Hydrocel system with 128 channels and a 500 Hz sampling rate. The impedance of each electrode was analyzed between recording sessions and kept at a maximum of 40 KOhm. Eye positions were concurrently recorded with an EyeLink 1000 Plus at the same sampling rate. The eye tracking was calibrated using a 9-point grid before each recording and validated to ensure an average error of less than $1^{\circ}$ for the measurement of all points. Participants were seated 68 cm from a 24-inch monitor with a resolution of 800x600 pixels, with their heads stabilized in a chin rest position.

% KOhm --> kΩ???
%What is ET?

To address artifacts present in the EEG recordings as a result of environmental and psychological noise, the necessary preprocessing steps were taken. This process involved the detection and correction of bad electrodes, along with running a 40 Hz high-pass filter and a 0.5 Hz low-pass filter on the data. The EEG data was then synchronized alongside the eye-tracking data to ensure time-locked analyses at the onset of relevant events with errors not exceeding 2 ms. 

In the Large Grid Paradigm, participants were asked to focus on a sequence of 25 dots located in different positions on the screen. Dots were each presented for roughly 1.5 to 1.8 seconds, and their positions were selected to ensure maximal coverage of the screen area. This procedure was separated into five experimental blocks, each displaying a series of 27 dots, with the center dot appearing three times in a pseudo-randomized order to reduce the predictability of subsequent dots. This entire experiment was then repeated six times.

The Absolute Position benchmark data was performed using the Large Grid Paradigm. This task involves determining the subject's gaze position as an XY coordinate pair. Each sample of one second describes a single fixation from a participant. The benchmark contains 21464 samples from 27 participants. This task was specifically important for our research as it provided a diverse array of gaze positions combined with a high sample count, allowing for a more comprehensive analysis of EEG-ET patterns, integral to determining the XY coordinate positions \citep{eegeyenet}.

%Expand more and define some of the key terminology to better explain either what it is, why we included it, and its effect/benefit

\begin{figure}[hbtp]
  \centering
  \includegraphics[width=1.0 \linewidth]{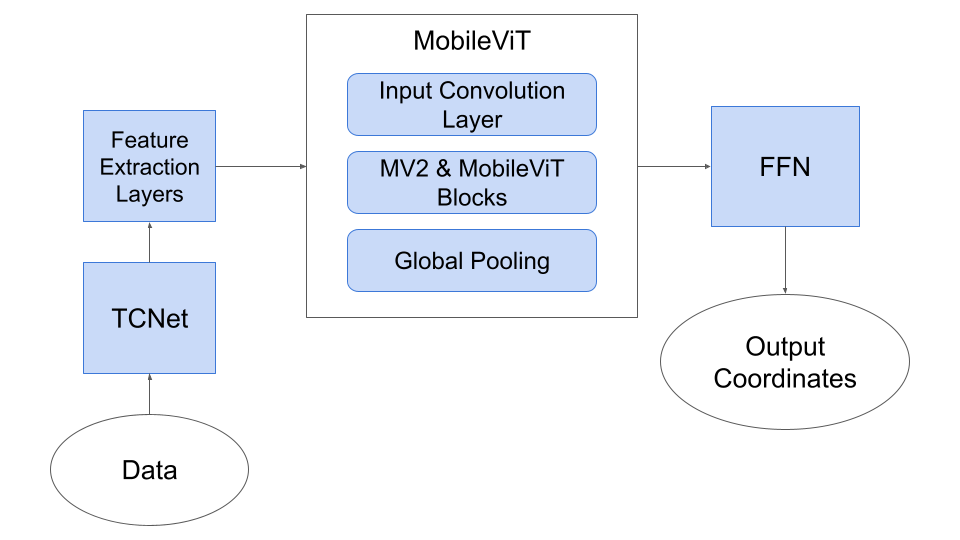}
  \caption{\textit{EEGMobile Model Flow Chart: MV2 refers to MobileNetV2 and FFN refers to Feed Forward Network.}}
  \label{fig:eegmobile}
\end{figure}

\subsection{3.2 Model Architecture}

The architecture of our model can be seen in Figure \ref{fig:eegmobile}. Our model design is adapted from the EEGViT-TCNet architecture and includes three main components that construct the entire network architecture.

\textbf{Temporal Convolutional Network Block:} The TCN block is initialized with a 129-dimension input layer, specifically attuned to the number of EEG channels with an extra dimension for encoding grounding information. Three additional layers with channel sizes 64, 128, and 256 are utilized to build a comprehensive representation of the temporal dependencies within the recorded EEG signals. A uniform kernel of size 3 coupled with ReLU activation is applied to all layers with a dropout rate of 0.75 to counteract overfitting. Finally, causality and weight normalization are applied to improve stability \cite{tcnet}.

\textbf{Feature Extraction Layers:} This block consists of two convolutional layers designed to initiate feature extraction on the data output by TCN in preparation to be fed to the MobileViT layer. The first layer consists of a convolutional kernel of size (1, 36) scaling the input up to 256 channels, initiating spatial feature extraction from the input data. The second layer consists of a kernel of size (256, 1), further scaling the data to 768 channels and compressing spatial information in preparation for the transformer component \cite{vit2eeg}.

\textbf{MobileViT Network:} This final block consists of a pre-trained MobileViTV2 network configured from the Hugging Face "apple/mobilevitv2-1.0-imagenet1k-256" model. The model is initialized with an input dimensionality of 768, perfectly aligning with the previous feature extraction component, along with a kernel size of (3, 3) for the internal convolutional layers and a patch size of (1, 1) for the internal transformers. Lastly, a classifier head is comprised of a classifier layer followed by a dropout of 0.1, finally connecting to a linear layer that outputs the final XY coordinate gaze prediction \cite{mobilevit, mobilevitv2, vit2eeg}.

\subsection{3.3 Training and Evaluation:}
Data from the Absolute Position tasks was split 70\% for training, 15\% for validation, and 15\% for testing. To maintain data integrity, we ensured data points with eye positions outside of the 800-by-600 pixel range were excluded. Our model was then trained for a total of 15 epochs, with each training iteration consisting of 64 sample batches \cite{eegeyenet}. Our model utilized the Adam optimizer with a learning rate of 1e-3 and weight decay of 0.3, and a learning rate scheduler with a step size of 6 was utilized to prevent overfitting. Additionally, we employed knowledge distillation using a fine-tuned EEGViT-TCN teacher model to train our EEGMobile student model, due to its superior performance over other models on the Absolute Position Task. The loss was calculated as a function of a lambda-controlled, weighted sum between the Mean Square Error (MSE) between the student predictions and observed values, the "true loss", and the Kullback-Leibler divergence loss of the soft targets of both models, the "distillation loss". We set the temperature parameter, controlling the strength of the distillation, to 20 and lambda to 0.9, giving a higher weight to the distillation loss. \cite{knowledgedistilling}.
%Mean Square Error =? RMSE

Model accuracy was evaluated as the Euclidean distance between the model's predictions and observed values in millimeters, represented as the RMSE \cite{eegeyenet}. We also evaluated the model's speed through an inference test, where the model was set to run inferencing across the entire Absolute Position task dataset ten times, after which the runtime in minutes was computed. Finally, we report parameter counts for the compared models to examine their relative sizes. All training and testing were conducted on the same machine using a single P5000 GPU.

\section{4. Results}%split up results and discussion sections
Results from our exhaustive evaluation illustrate EEGMobile's ability to deliver SOTA-comparable accuracy at a faster inferencing speed and smaller size, outperforming the other transformer-based models.

\begin{table}[hbtp]

\caption{\textit{RMSE indicates validation error in millimeters. Runtime measures the time to complete inferencing on the entire dataset 10 times. Parameter count is measured in millions. All values represent the mean and standard deviation of 5 runs. All tests were run on the same hardware.}}
\label{tab:model_speed}
\resizebox{1\textwidth}{!}{%
\centering
\begin{tabular}{|l|c|c|c|}
\hline
\textbf{Model} & \textbf{RMSE (mm)} & \textbf{Runtime (mins)} & \textbf{Parameter Count}\\ \hline
CNN \cite{eegeyenet} & 70.4 ± 1.1 & 2.1 ± 0.2 & 0.6 M\\ \hline
EEGViT \cite{vit2eeg} & 55.4 ± 0.2 & 78.8 ± 0.2 & 86.0 M\\ \hline
EEGViT-TCN \cite{modesitt2024fusing} & 51.8 ± 0.2 & 12.1 ± 1.6 & 137.2 M\\ \hline \hline
\textbf{EEGMobile (Ours)} & \textbf{53.6 ± 0.6} & \textbf{8.1 ± 0.9} & \textbf{55.9 M}\\ \hline
\end{tabular}

}
\end{table}

Table \ref{tab:model_speed} shows that our EEGMobile architecture attains an impressively low RMSE of 53.6 mm. This eclipses all EEGEyeNet baseline models, the lowest of which is the CNN with an RMSE of 70.4 \citep{eegeyenet}. 
Likewise, our model outperforms the highest-performing EEGViT model with an RMSE of 55.4 \citep{vit2eeg}. EEGMobile also attains a similar accuracy to EEGViT-TCNet, just 3\% shy of its 51.8 RMSE \citep{modesitt2024fusing}. This notable performance accuracy underscores the robustness of EEGMobile and its ability to capture and generalize key information within the training data.

In addition to evaluating the accuracy, we also assessed EEGMobile's computational efficiency through an inference speed test and analyzed the model's relative size. Compared to the other EEG-based ViT architectures, our model boasts a 33\% increase in inferencing speed when compared to EEGViT-TCNet and is over 9.7$\times$ faster than the original EEGViT with a runtime of about 8.1 minutes, compared to 78.8 and 12.1 minutes for EEGViT and EEGViT-TCNet, respectively. However, EEGMobile is still slower than the baseline CNN model by about 3.9$\times$. Furthermore, we find that while EEGMobile has a smaller size compared to the other ViT architectures, with 55.9 million parameters, followed by EEGViT with 86 million and EEGViT-TCN with 137.2 million, It is still much larger than the CNN model with only 609 thousand parameters. These sizes indicate that while EEGMobile is still about 91.7$\times$ larger than the CNN, it is also about 35\% smaller than EEGViT and 60\% smaller than EEGViT-TCN. This noteworthy improvement to both speed and size highlights EEGMobile's efficiency over the other transformer-based models and is especially paramount for applications involving real-time predictions on memory-restricted devices.

\section{5. Discussion} 
Our meticulous performance evaluation legitimizes EEGMobile as a robust model capable of executing EEG-ET tasks efficiently when compared to other transformer-based models, reinforcing the viability of combining lightweight networks with distillation techniques to develop cost-effective models. Our RMSE results suggest that EEGMobile is capable of providing predictive capabilities comparable to the other transformer models on EEG gaze estimation tasks and can gain a robust interpretation of the EEG data despite its small size. Similarly, without the use of KD, EEGMobile has a much higher RMSE of about 76.8. This alludes to the idea that KD aids in guiding the model to learn better representations, enhancing its accuracy. This is further highlighted by the high temperature and lambda parameters, which greatly increased the strength and effect of distillation. We also note that EEGMobile's memory and time efficiency are still very much below those of CNN. However, when examining the components of all three ViT-based models, a significant amount of the total parameter counts stem from the feature extraction layers, which contain over 50 million parameters. As such, there is a strong possibility that integrating feature extraction methods with significantly fewer parameters may further close the size gap between EEGMobile and CNN in terms of speed and size. Our analysis of EEGMobile's size, speed, and task performance illustrates its potential as an efficient model for practical EEG signal analysis, especially on weaker devices, across various domains.

This improved efficiency is material to real-time EEG and eye-tracking applications, where speed and accuracy are important \cite{ELMENSHAWY2015mobileseizure}. Our findings are also particularly relevant to the HCI community, such as in the integration of EEG recording devices into virtual reality and augmented reality technologies to improve interactivity \cite{xiang2022vector, rolff2022gazetransformer, xu2023cmfs}. Additionally, due to the model's low parameter count, there are also memory efficiency benefits when it is utilized by resource-constrained BCI devices, allowing the model to be run on-device, making medical diagnostic tools more accessible to the average person \cite{jebelli2019eegstess}.

Due to time constraints, our experiment only tested the MobileViT and MobileViTV2 architectures, ultimately selecting the V2 in our final results. However, a MobileViTV3, which makes a number of improvements to further enhance the model's performance, has also been developed \cite{wadekar2022mobilevitv3}. Not only does utilizing this pre-trained model have the potential to improve our results, but it would also be more representative of the capabilities of MobileViT on the Absolute Position Task. Similarly, testing our model on other EEG datasets would allow us to better evaluate its robustness for EEG analysis as a whole.

In future studies, exploring alternative lightweight model architectures has the potential to yield better accuracy and speed. Experimenting with alternative distillation techniques, such as feature distillation, may also lead to improved performance. Additionally, since there is no theoretical constraint on the teacher model used in the distillation paradigm, future research could focus on developing a large, high-performance model to train a smaller, more practical model using the same training procedure as EEGMobile. One notable concern with the current methodology is the independent training of both the teacher and student models, which can be inefficient, especially with very large and slow teacher models. Research into parallelizing this training process may increase the scalability of these models. Furthermore, exploring various deep learning techniques on different datasets for comparative studies (\cite{an2023transfer,an2023survey,jiang2023successfully,lu2023machine,chen2024trialbench,ma2022traffic,ma2024data,gui2024remote,tan2023audio,tan2021multivariate,qiu2023modal,zhao2024deep,zhang2022attention,zhang2023trep}) could provide valuable insights and further enhance the field.

%too long make it smaller
\section{6. Conclusion}
This study highlights the efficacy of lightweight models for EEG-ET tasks. By integrating the MobileViT architecture into a Hybrid Transformer model and utilizing knowledge distillation techniques, we present a model with enhanced speed and size with a minimal cost to accuracy compared to the SOTA. Our findings further validate the potential of lightweight networks on EEG regression tasks, but further expand the accessibility of DL-based EEG analysis tools for real-world applications, especially on resource-constrained devices, with the potential to advance the unification of neuroscience and HCI.

\bibliography{EEGMobile}

\begin{thebibliography}{68}
\providecommand{\natexlab}[1]{#1}
\providecommand{\url}[1]{\texttt{#1}}
\expandafter\ifx\csname urlstyle\endcsname\relax
  \providecommand{\doi}[1]{doi: #1}\else
  \providecommand{\doi}{doi: \begingroup \urlstyle{rm}\Url}\fi

\bibitem[Aggarwal and Chugh(2022)]{aggarwal2022review}
Swati Aggarwal and Nupur Chugh.
\newblock Review of machine learning techniques for eeg based brain computer interface.
\newblock \emph{Archives of Computational Methods in Engineering}, 29\penalty0 (5):\penalty0 3001--3020, 2022.

\bibitem[Altaheri et~al.(2023)Altaheri, Muhammad, Alsulaiman, Amin, Altuwaijri, Abdul, Bencherif, and Faisal]{altaheri2023deep}
Hamdi Altaheri, Ghulam Muhammad, Mansour Alsulaiman, Syed~Umar Amin, Ghadir~Ali Altuwaijri, Wadood Abdul, Mohamed~A Bencherif, and Mohammed Faisal.
\newblock Deep learning techniques for classification of electroencephalogram (eeg) motor imagery (mi) signals: A review.
\newblock \emph{Neural Computing and Applications}, 35\penalty0 (20):\penalty0 14681--14722, 2023.

\bibitem[An et~al.(2023{\natexlab{a}})An, Bhat, Gumussoy, and Ogras]{an2023transfer}
Sizhe An, Ganapati Bhat, Suat Gumussoy, and Umit Ogras.
\newblock Transfer learning for human activity recognition using representational analysis of neural networks.
\newblock \emph{ACM Transactions on Computing for Healthcare}, 4\penalty0 (1):\penalty0 1--21, 2023{\natexlab{a}}.

\bibitem[An et~al.(2023{\natexlab{b}})An, Tuncel, Basaklar, and Ogras]{an2023survey}
Sizhe An, Yigit Tuncel, Toygun Basaklar, and Umit~Y Ogras.
\newblock A survey of embedded machine learning for smart and sustainable healthcare applications.
\newblock In \emph{Embedded Machine Learning for Cyber-Physical, IoT, and Edge Computing: Use Cases and Emerging Challenges}, pages 127--150. Springer, 2023{\natexlab{b}}.

\bibitem[Bai et~al.(2018)Bai, Kolter, and Koltun]{tcnet}
Shaojie Bai, J.~Zico Kolter, and Vladlen Koltun.
\newblock An empirical evaluation of generic convolutional and recurrent networks for sequence modeling.
\newblock \emph{arXiv preprint arXiv:1803.01271}, 2018.

\bibitem[Cao(2020)]{cao2020eegai}
Zehong Cao.
\newblock A review of artificial intelligence for eeg-based brain-computer interfaces and applications.
\newblock \emph{Brain Science Advances}, 6\penalty0 (3):\penalty0 162--170, 2020.
\newblock \doi{10.26599/BSA.2020.9050017}.
\newblock URL \url{https://doi.org/10.26599/BSA.2020.9050017}.

\bibitem[Chen et~al.(2024)Chen, Hu, Wang, Lu, Cao, Lin, Xu, Wu, Xiao, Sun, et~al.]{chen2024trialbench}
Jintai Chen, Yaojun Hu, Yue Wang, Yingzhou Lu, Xu~Cao, Miao Lin, Hongxia Xu, Jian Wu, Cao Xiao, Jimeng Sun, et~al.
\newblock Trialbench: Multi-modal artificial intelligence-ready clinical trial datasets.
\newblock \emph{arXiv preprint arXiv:2407.00631}, 2024.

\bibitem[Craik et~al.(2019)Craik, He, and Contreras-Vidal]{craik2019deep}
Alexander Craik, Yongtian He, and Jose~L Contreras-Vidal.
\newblock Deep learning for electroencephalogram (eeg) classification tasks: a review.
\newblock \emph{Journal of neural engineering}, 16\penalty0 (3):\penalty0 031001, 2019.

\bibitem[Cui et~al.(2024)Cui, Guo, Cao, Tang, Wen, Jin, Wang, and Hou]{cui2024lightweight}
Yiming Cui, Jiajia Guo, Zheng Cao, Huaze Tang, Chao-Kai Wen, Shi Jin, Xin Wang, and Xiaolin Hou.
\newblock Lightweight neural network with knowledge distillation for csi feedback, 2024.

\bibitem[Dadebayev et~al.(2022)Dadebayev, Goh, and Tan]{dadebayev2022eeg}
Didar Dadebayev, Wei~Wei Goh, and Ee~Xion Tan.
\newblock Eeg-based emotion recognition: Review of commercial eeg devices and machine learning techniques.
\newblock \emph{Journal of King Saud University-Computer and Information Sciences}, 34\penalty0 (7):\penalty0 4385--4401, 2022.

\bibitem[Dosovitskiy et~al.(2021)Dosovitskiy, Beyer, Kolesnikov, Weissenborn, Zhai, Unterthiner, Dehghani, Minderer, Heigold, Gelly, Uszkoreit, and Houlsby]{dosovitskiy2021image}
Alexey Dosovitskiy, Lucas Beyer, Alexander Kolesnikov, Dirk Weissenborn, Xiaohua Zhai, Thomas Unterthiner, Mostafa Dehghani, Matthias Minderer, Georg Heigold, Sylvain Gelly, Jakob Uszkoreit, and Neil Houlsby.
\newblock An image is worth 16x16 words: Transformers for image recognition at scale, 2021.

\bibitem[Dou et~al.(2022)Dou, Zhou, and Qu]{dou2022time}
Guangyao Dou, Zheng Zhou, and Xiaodong Qu.
\newblock Time majority voting, a pc-based eeg classifier for non-expert users.
\newblock In \emph{International Conference on Human-Computer Interaction}, pages 415--428. Springer, 2022.

\bibitem[{EL Menshawy} et~al.(2015){EL Menshawy}, Benharref, and Serhani]{ELMENSHAWY2015mobileseizure}
Mohamed {EL Menshawy}, Abdelghani Benharref, and Mohamed Serhani.
\newblock An automatic mobile-health based approach for eeg epileptic seizures detection.
\newblock \emph{Expert Systems with Applications}, 42\penalty0 (20):\penalty0 7157--7174, 2015.
\newblock ISSN 0957-4174.
\newblock \doi{https://doi.org/10.1016/j.eswa.2015.04.068}.
\newblock URL \url{https://www.sciencedirect.com/science/article/pii/S0957417415003103}.

\bibitem[Farago et~al.(2022)Farago, Law, Hajra, and Chan]{farago2022blink}
Emma Farago, Andrew~J Law, Sujoy~Ghosh Hajra, and Adrian~DC Chan.
\newblock Blink and saccade detection from forehead eeg.
\newblock In \emph{2022 IEEE International Instrumentation and Measurement Technology Conference (I2MTC)}, pages 1--6. IEEE, 2022.

\bibitem[Fuhl et~al.(2023)Fuhl, Zabel, Harbig, Moldt, Wietek, Herrmann-Werner, and Nieselt]{fuhl2023one}
Wolfgang Fuhl, Susanne Zabel, Theresa Harbig, Julia-Astrid Moldt, Teresa~Festl Wietek, Anne Herrmann-Werner, and Kay Nieselt.
\newblock One step closer to eeg based eye tracking.
\newblock In \emph{Proceedings of the 2023 Symposium on Eye Tracking Research and Applications}, pages 1--7, 2023.

\bibitem[Gao and Zhou(2023)]{gao2023sisr}
Dandan Gao and Dengwen Zhou.
\newblock A very lightweight and efficient image super-resolution network.
\newblock \emph{Expert Systems with Applications}, 213:\penalty0 118898, 2023.
\newblock ISSN 0957-4174.
\newblock \doi{https://doi.org/10.1016/j.eswa.2022.118898}.
\newblock URL \url{https://www.sciencedirect.com/science/article/pii/S0957417422019169}.

\bibitem[Gao et~al.(2021)Gao, Dang, Wang, Hong, Hou, Ma, and Perc]{gao2021complex}
Zhongke Gao, Weidong Dang, Xinmin Wang, Xiaolin Hong, Linhua Hou, Kai Ma, and Matja{\v{z}} Perc.
\newblock Complex networks and deep learning for eeg signal analysis.
\newblock \emph{Cognitive Neurodynamics}, 15\penalty0 (3):\penalty0 369--388, 2021.

\bibitem[Gui et~al.(2024)Gui, Song, Qin, and Tang]{gui2024remote}
Shengxi Gui, Shuang Song, Rongjun Qin, and Yang Tang.
\newblock Remote sensing object detection in the deep learning era—a review.
\newblock \emph{Remote Sensing}, 16\penalty0 (2):\penalty0 327, 2024.

\bibitem[Hinton et~al.(2015)Hinton, Vinyals, and Dean]{knowledgedistilling}
Geoffrey Hinton, Oriol Vinyals, and Jeff Dean.
\newblock Distilling the knowledge in a neural network, 2015.

\bibitem[Hossain et~al.(2023)Hossain, Islam, Hossain, Nijholt, and Ahad]{hossain2023status}
Khondoker~Murad Hossain, Md~Ariful Islam, Shahera Hossain, Anton Nijholt, and Md~Atiqur~Rahman Ahad.
\newblock Status of deep learning for eeg-based brain--computer interface applications.
\newblock \emph{Frontiers in computational neuroscience}, 16:\penalty0 1006763, 2023.

\bibitem[Ingolfsson et~al.(2020)Ingolfsson, Hersche, Wang, Kobayashi, Cavigelli, and Benini]{eegtcn}
Thorir~Mar Ingolfsson, Michael Hersche, Xiaying Wang, Nobuaki Kobayashi, Lukas Cavigelli, and Luca Benini.
\newblock Eeg-tcnet: An accurate temporal convolutional network for embedded motor-imagery brain–machine interfaces.
\newblock In \emph{2020 IEEE International Conference on Systems, Man, and Cybernetics (SMC)}, pages 2958--2965, 2020.
\newblock \doi{10.1109/SMC42975.2020.9283028}.

\bibitem[Jebelli et~al.(2019)Jebelli, Khalili, and Lee]{jebelli2019eegstess}
Houtan Jebelli, Mohammad~Mahdi Khalili, and SangHyun Lee.
\newblock Mobile eeg-based workers' stress recognition by applying deep neural network.
\newblock In Ivan Mutis and Timo Hartmann, editors, \emph{Advances in Informatics and Computing in Civil and Construction Engineering}, pages 173--180, Cham, 2019. Springer International Publishing.
\newblock ISBN 978-3-030-00220-6.

\bibitem[Jiang et~al.(2023)Jiang, Hui, Liu, and Yan]{jiang2023successfully}
Chao Jiang, Bo~Hui, Bohan Liu, and Da~Yan.
\newblock Successfully applying lottery ticket hypothesis to diffusion model.
\newblock \emph{arXiv preprint arXiv:2310.18823}, 2023.

\bibitem[Kastrati et~al.(2021{\natexlab{a}})Kastrati, P{\l}omecka, Pascual, Wolf, Gillioz, Wattenhofer, and Langer]{kastrati2021eegeyenet}
Ard Kastrati, Martyna~Beata P{\l}omecka, Dami{\'a}n Pascual, Lukas Wolf, Victor Gillioz, Roger Wattenhofer, and Nicolas Langer.
\newblock Eegeyenet: a simultaneous electroencephalography and eye-tracking dataset and benchmark for eye movement prediction.
\newblock \emph{arXiv preprint arXiv:2111.05100}, 2021{\natexlab{a}}.

\bibitem[Kastrati et~al.(2021{\natexlab{b}})Kastrati, Płomecka, Pascual, Wolf, Gillioz, Wattenhofer, and Langer]{eegeyenet}
Ard Kastrati, Martyna~Beata Płomecka, Damián Pascual, Lukas Wolf, Victor Gillioz, Roger Wattenhofer, and Nicolas Langer.
\newblock Eegeyenet: a simultaneous electroencephalography and eye-tracking dataset and benchmark for eye movement prediction, 2021{\natexlab{b}}.

\bibitem[Kastrati et~al.(2023)Kastrati, Plomecka, K{\"u}chler, Langer, and Wattenhofer]{kastrati2023electrode}
Ard Kastrati, Martyna~Beata Plomecka, Jo{\"e}l K{\"u}chler, Nicolas Langer, and Roger Wattenhofer.
\newblock Electrode clustering and bandpass analysis of eeg data for gaze estimation.
\newblock In \emph{Annual Conference on Neural Information Processing Systems}, pages 50--65. PMLR, 2023.

\bibitem[Key et~al.(2024)Key, Mehtiyev, and Qu]{key2024advancing}
Matthew~L Key, Tural Mehtiyev, and Xiaodong Qu.
\newblock Advancing eeg-based gaze prediction using depthwise separable convolution and enhanced pre-processing.
\newblock In \emph{International Conference on Human-Computer Interaction}, pages 3--17. Springer, 2024.

\bibitem[Koome~Murungi et~al.(2023)Koome~Murungi, Pham, Dai, and Qu]{koome2023trends}
Nathan Koome~Murungi, Michael~Vinh Pham, Xufeng Dai, and Xiaodong Qu.
\newblock Trends in machine learning and electroencephalogram (eeg): A review for undergraduate researchers.
\newblock \emph{arXiv e-prints}, pages arXiv--2307, 2023.

\bibitem[Li et~al.(2020)Li, Lee, Jung, Youn, and Camacho]{li2020deep}
Gen Li, Chang~Ha Lee, Jason~J Jung, Young~Chul Youn, and David Camacho.
\newblock Deep learning for eeg data analytics: A survey.
\newblock \emph{Concurrency and Computation: Practice and Experience}, 32\penalty0 (18):\penalty0 e5199, 2020.

\bibitem[Li et~al.(2024)Li, Zhou, and Qu]{li2024enhancing}
Weigeng Li, Neng Zhou, and Xiaodong Qu.
\newblock Enhancing eye-tracking performance through multi-task learning transformer.
\newblock In \emph{International Conference on Human-Computer Interaction}, pages 31--46. Springer, 2024.

\bibitem[Lu et~al.(2023)Lu, Shen, Wang, Wang, van Rechem, and Wei]{lu2023machine}
Yingzhou Lu, Minjie Shen, Huazheng Wang, Xiao Wang, Capucine van Rechem, and Wenqi Wei.
\newblock Machine learning for synthetic data generation: a review.
\newblock \emph{arXiv preprint arXiv:2302.04062}, 2023.

\bibitem[Ma(2022)]{ma2022traffic}
Xiaobo Ma.
\newblock \emph{Traffic performance evaluation using statistical and machine learning methods}.
\newblock PhD thesis, The University of Arizona, 2022.

\bibitem[Ma et~al.(2024)Ma, Karimpour, and Wu]{ma2024data}
Xiaobo Ma, Abolfazl Karimpour, and Yao-Jan Wu.
\newblock Data-driven transfer learning framework for estimating on-ramp and off-ramp traffic flows.
\newblock \emph{Journal of Intelligent Transportation Systems}, pages 1--14, 2024.

\bibitem[Mehta and Rastegari(2022{\natexlab{a}})]{mobilevit}
Sachin Mehta and Mohammad Rastegari.
\newblock Mobilevit: Light-weight, general-purpose, and mobile-friendly vision transformer, 2022{\natexlab{a}}.

\bibitem[Mehta and Rastegari(2022{\natexlab{b}})]{mobilevitv2}
Sachin Mehta and Mohammad Rastegari.
\newblock Separable self-attention for mobile vision transformers, 2022{\natexlab{b}}.

\bibitem[Mishra et~al.(2023)Mishra, Kumar, Gupta, Prabhu, Upadhyay, Chhipa, Rakesh, Mokayed, Liwicki, Liwicki, et~al.]{mishra2023signeeg}
Ashish~Ranjan Mishra, Rakesh Kumar, Vibha Gupta, Sameer Prabhu, Richa Upadhyay, Prakash~Chandra Chhipa, Sumit Rakesh, Hamam Mokayed, Marcus Liwicki, Foteini~Simistira Liwicki, et~al.
\newblock Signeeg v1. 0: Multimodal electroencephalography and signature database for biometric systems.
\newblock \emph{bioRxiv}, pages 2023--09, 2023.

\bibitem[Modesitt et~al.(2023)Modesitt, Yang, and Liu]{modesitt2023two}
Eric Modesitt, Ruiqi Yang, and Qi~Liu.
\newblock Two heads are better than one: A bio-inspired method for improving classification on eeg-et data.
\newblock In \emph{International Conference on Human-Computer Interaction}, pages 382--390. Springer, 2023.

\bibitem[Modesitt et~al.(2024)Modesitt, Yin, Wang, and Lu]{modesitt2024fusing}
Eric Modesitt, Haicheng Yin, Williams~Huang Wang, and Brian Lu.
\newblock Fusing pretrained vits with tcnet for enhanced eeg regression, 2024.

\bibitem[Murungi et~al.(2023{\natexlab{a}})Murungi, Pham, Dai, and Qu]{murungi2023trends}
Nathan~Koome Murungi, Michael~Vinh Pham, Xufeng Dai, and Xiaodong Qu.
\newblock Trends in machine learning and electroencephalogram (eeg): A review for undergraduate researchers.
\newblock In \emph{International Conference on Human-Computer Interaction}, pages 426--443. Springer, 2023{\natexlab{a}}.

\bibitem[Murungi et~al.(2023{\natexlab{b}})Murungi, Pham, Dai, and Qu]{murungi2023empowering}
Nathan~Koome Murungi, Michael~Vinh Pham, Xufeng~Caesar Dai, and Xiaodong Qu.
\newblock Empowering computer science students in electroencephalography (eeg) analysis: A review of machine learning algorithms for eeg datasets.
\newblock \emph{SIGKDD}, 2023{\natexlab{b}}.

\bibitem[Qiu et~al.(2023)Qiu, Zhao, Yao, Chen, and Wang]{qiu2023modal}
Yansheng Qiu, Ziyuan Zhao, Hongdou Yao, Delin Chen, and Zheng Wang.
\newblock Modal-aware visual prompting for incomplete multi-modal brain tumor segmentation.
\newblock In \emph{Proceedings of the 31st ACM International Conference on Multimedia}, pages 3228--3239, 2023.

\bibitem[Qu(2022)]{qu2022time}
Xiaodong Qu.
\newblock \emph{Time Continuity Voting for Electroencephalography (EEG) Classification}.
\newblock PhD thesis, Brandeis University, 2022.

\bibitem[Qu and Hickey(2022)]{qu2022eeg4home}
Xiaodong Qu and Timothy~J Hickey.
\newblock Eeg4home: A human-in-the-loop machine learning model for eeg-based bci.
\newblock In \emph{Augmented Cognition: 16th International Conference, AC 2022, Held as Part of the 24th HCI International Conference, HCII 2022, Virtual Event, June 26--July 1, 2022, Proceedings}, pages 162--172. Springer, 2022.

\bibitem[Qu et~al.(2018)Qu, Sun, Sekuler, and Hickey]{qu2018eeg}
Xiaodong Qu, Yixin Sun, Robert Sekuler, and Timothy Hickey.
\newblock Eeg markers of stem learning.
\newblock In \emph{2018 IEEE Frontiers in Education Conference (FIE)}, pages 1--9. IEEE, 2018.

\bibitem[Qu et~al.(2019)Qu, Hall, Sun, Sekuler, and Hickey]{qu2019personalized}
Xiaodong Qu, Mercedes Hall, Yile Sun, Robert Sekuler, and Timothy~J Hickey.
\newblock A personalized reading coach using wearable eeg sensors.
\newblock \emph{CSEDU}, 2019.

\bibitem[Qu et~al.(2020{\natexlab{a}})Qu, Liu, Li, and Hickey]{qu2020multi}
Xiaodong Qu, Peiyan Liu, Zhaonan Li, and Timothy Hickey.
\newblock Multi-class time continuity voting for eeg classification.
\newblock In \emph{Brain Function Assessment in Learning: Second International Conference, BFAL 2020, Heraklion, Crete, Greece, October 9--11, 2020, Proceedings 2}, pages 24--33. Springer, 2020{\natexlab{a}}.

\bibitem[Qu et~al.(2020{\natexlab{b}})Qu, Liukasemsarn, Tu, Higgins, Hickey, and Hall]{qu2020identifying}
Xiaodong Qu, Saran Liukasemsarn, Jingxuan Tu, Amy Higgins, Timothy~J Hickey, and Mei-Hua Hall.
\newblock Identifying clinically and functionally distinct groups among healthy controls and first episode psychosis patients by clustering on eeg patterns.
\newblock \emph{Frontiers in psychiatry}, 11:\penalty0 541659, 2020{\natexlab{b}}.

\bibitem[Qu et~al.(2020{\natexlab{c}})Qu, Mei, Liu, and Hickey]{qu2020using}
Xiaodong Qu, Qingtian Mei, Peiyan Liu, and Timothy Hickey.
\newblock Using eeg to distinguish between writing and typing for the same cognitive task.
\newblock In \emph{Brain Function Assessment in Learning: Second International Conference, BFAL 2020, Heraklion, Crete, Greece, October 9--11, 2020, Proceedings 2}, pages 66--74. Springer, 2020{\natexlab{c}}.

\bibitem[Rasheed et~al.(2020)Rasheed, Qayyum, Qadir, Sivathamboo, Kwan, Kuhlmann, O’Brien, and Razi]{rasheed2020machine}
Khansa Rasheed, Adnan Qayyum, Junaid Qadir, Shobi Sivathamboo, Patrick Kwan, Levin Kuhlmann, Terence O’Brien, and Adeel Razi.
\newblock Machine learning for predicting epileptic seizures using eeg signals: A review.
\newblock \emph{IEEE reviews in biomedical engineering}, 14:\penalty0 139--155, 2020.

\bibitem[Rolff et~al.(2022)Rolff, Harms, Steinicke, and Frintrop]{rolff2022gazetransformer}
Tim Rolff, H~Matthias Harms, Frank Steinicke, and Simone Frintrop.
\newblock Gazetransformer: Gaze forecasting for virtual reality using transformer networks.
\newblock In \emph{DAGM German Conference on Pattern Recognition}, pages 577--593. Springer, 2022.

\bibitem[Roy et~al.(2019)Roy, Banville, Albuquerque, Gramfort, Falk, and Faubert]{roy2019deep}
Yannick Roy, Hubert Banville, Isabela Albuquerque, Alexandre Gramfort, Tiago~H Falk, and Jocelyn Faubert.
\newblock Deep learning-based electroencephalography analysis: a systematic review.
\newblock \emph{Journal of neural engineering}, 16\penalty0 (5):\penalty0 051001, 2019.

\bibitem[Saeidi et~al.(2021)Saeidi, Karwowski, Farahani, Fiok, Taiar, Hancock, and Al-Juaid]{saeidi2021neural}
Maham Saeidi, Waldemar Karwowski, Farzad~V Farahani, Krzysztof Fiok, Redha Taiar, Peter~A Hancock, and Awad Al-Juaid.
\newblock Neural decoding of eeg signals with machine learning: a systematic review.
\newblock \emph{Brain Sciences}, 11\penalty0 (11):\penalty0 1525, 2021.

\bibitem[Sun and Mou(2023)]{eegdirection}
Congzhong Sun and Chaozhou Mou.
\newblock Survey on the research direction of eeg-based signal processing.
\newblock \emph{Frontiers in Neuroscience}, 17, July 2023.
\newblock ISSN 1662-453X.
\newblock \doi{10.3389/fnins.2023.1203059}.
\newblock URL \url{http://dx.doi.org/10.3389/fnins.2023.1203059}.

\bibitem[Tan et~al.(2021)Tan, Shen, Zhang, and Wang]{tan2021multivariate}
Jieyuan Tan, Xiang Shen, Xiang Zhang, and Yiwen Wang.
\newblock Multivariate encoding analysis of medial prefrontal cortex cortical activity during task learning.
\newblock In \emph{2021 43rd Annual International Conference of the IEEE Engineering in Medicine \& Biology Society (EMBC)}, pages 6699--6702. IEEE, 2021.

\bibitem[Tan et~al.(2023)Tan, Zhang, Wu, Song, Chen, Huang, and Wang]{tan2023audio}
Jieyuan Tan, Xiang Zhang, Shenghui Wu, Zhiwei Song, Shuhang Chen, Yifan Huang, and Yiwen Wang.
\newblock Audio-induced medial prefrontal cortical dynamics enhances coadaptive learning in brain--machine interfaces.
\newblock \emph{Journal of Neural Engineering}, 20\penalty0 (5):\penalty0 056035, 2023.

\bibitem[Wadekar and Chaurasia(2022)]{wadekar2022mobilevitv3}
Shakti~N. Wadekar and Abhishek Chaurasia.
\newblock Mobilevitv3: Mobile-friendly vision transformer with simple and effective fusion of local, global and input features, 2022.

\bibitem[Wang and Qu(2022)]{wang2022eeg}
Ruyang Wang and Xiaodong Qu.
\newblock Eeg daydreaming, a machine learning approach to detect daydreaming activities.
\newblock In \emph{International Conference on Human-Computer Interaction}, pages 202--212. Springer, 2022.

\bibitem[Wang and Wang(2022)]{wang2022cnn}
Xuduo Wang and Ziji Wang.
\newblock Cnn with self-attention in eeg classification.
\newblock In \emph{International Conference on Human-Computer Interaction}, pages 512--526. Springer, 2022.

\bibitem[Wolf et~al.(2022)Wolf, Kastrati, P{\l}omecka, Li, Klebe, Veicht, Wattenhofer, and Langer]{wolf2022deep}
Lukas Wolf, Ard Kastrati, Martyna~Beata P{\l}omecka, Jie-Ming Li, Dustin Klebe, Alexander Veicht, Roger Wattenhofer, and Nicolas Langer.
\newblock A deep learning approach for the segmentation of electroencephalography data in eye tracking applications.
\newblock \emph{arXiv preprint arXiv:2206.08672}, 2022.

\bibitem[Xiang and Abdelmonsef(2022)]{xiang2022vector}
Brian Xiang and Abdelrahman Abdelmonsef.
\newblock Vector-based data improves left-right eye-tracking classifier performance after a covariate distributional shift.
\newblock In \emph{International Conference on Human-Computer Interaction}, pages 617--632. Springer, 2022.

\bibitem[Xiao et~al.(2021)Xiao, Singh, Mintun, Darrell, Dollár, and Girshick]{xiao2021early}
Tete Xiao, Mannat Singh, Eric Mintun, Trevor Darrell, Piotr Dollár, and Ross Girshick.
\newblock Early convolutions help transformers see better, 2021.

\bibitem[Xu et~al.(2023)Xu, Yang, Yan, and Li]{xu2023cmfs}
Xu~Xu, Lei Yang, Yan Yan, and Congsheng Li.
\newblock Cmfs-net: Common mode features suppression network for gaze estimation.
\newblock In \emph{Proceedings of the 2023 Workshop on Advanced Multimedia Computing for Smart Manufacturing and Engineering}, AMC-SME '23, page 25–29, New York, NY, USA, 2023. Association for Computing Machinery.
\newblock ISBN 9798400702730.
\newblock \doi{10.1145/3606042.3616449}.
\newblock URL \url{https://doi.org/10.1145/3606042.3616449}.

\bibitem[Yang and Modesitt(2023)]{vit2eeg}
Ruiqi Yang and Eric Modesitt.
\newblock Vit2eeg: Leveraging hybrid pretrained vision transformers for eeg data, 2023.

\bibitem[Yi and Qu(2022)]{yi2022attention}
Long Yi and Xiaodong Qu.
\newblock Attention-based cnn capturing eeg recording’s average voltage and local change.
\newblock In \emph{Artificial Intelligence in HCI: 3rd International Conference, AI-HCI 2022, Held as Part of the 24th HCI International Conference, HCII 2022, Virtual Event, June 26--July 1, 2022, Proceedings}, pages 448--459. Springer, 2022.

\bibitem[Zhang et~al.(2022)Zhang, Tian, Sherony, Domeyer, and Ding]{zhang2022attention}
Zhengming Zhang, Renran Tian, Rini Sherony, Joshua Domeyer, and Zhengming Ding.
\newblock Attention-based interrelation modeling for explainable automated driving.
\newblock \emph{IEEE Transactions on Intelligent Vehicles}, 8\penalty0 (2):\penalty0 1564--1573, 2022.

\bibitem[Zhang et~al.(2023)Zhang, Tian, and Ding]{zhang2023trep}
Zhengming Zhang, Renran Tian, and Zhengming Ding.
\newblock Trep: Transformer-based evidential prediction for pedestrian intention with uncertainty.
\newblock In \emph{Proceedings of the AAAI Conference on Artificial Intelligence}, pages 3534--3542, 2023.

\bibitem[Zhao et~al.(2024)Zhao, Yang, Zeng, Qian, Zhao, Dai, Prabhu, Nordlund, and Tam]{zhao2024deep}
Shenghao Zhao, Xulei Yang, Zeng Zeng, Peisheng Qian, Ziyuan Zhao, Lingyun Dai, Nayana Prabhu, P{\"a}r Nordlund, and Wai~Leong Tam.
\newblock Deep learning based cetsa feature prediction cross multiple cell lines with latent space representation.
\newblock \emph{Scientific Reports}, 14\penalty0 (1):\penalty0 1878, 2024.

\bibitem[Zhou et~al.(2022)Zhou, Dou, and Qu]{zhou2022brainactivity1}
Zheng Zhou, Guangyao Dou, and Xiaodong Qu.
\newblock Brainactivity1: A framework of eeg data collection and machine learning analysis for college students.
\newblock In \emph{International Conference on Human-Computer Interaction}, pages 119--127. Springer, 2022.

\end{thebibliography}

\end{document}